\documentclass[prd,showpacs,superscriptaddress,nofootinbib,floatfix,10pt]{revtex4}
\usepackage{graphicx}
\usepackage{amsmath}
\usepackage{bm}
\usepackage{yhmath}
\usepackage{mathtools}
\usepackage{wasysym}
\usepackage[colorlinks,citecolor=violet,urlcolor=violet,linkcolor=blue]{hyperref}
\usepackage{subfigure}
\usepackage{color}
\usepackage{cases}
\usepackage{subfigure}
\usepackage{times}
\usepackage{dcolumn,booktabs,bm}
\usepackage{slashed}
\usepackage{amsfonts,amssymb,stmaryrd,latexsym,amsmath}
\usepackage{textcomp}
\usepackage{multirow}
\usepackage{cancel}
\usepackage{array}
\usepackage{orcidlink}
\usepackage{longtable}
\usepackage[utf8]{inputenc}
\usepackage[T1]{fontenc}
\usepackage[utf8]{inputenc}
\usepackage{cprotect}
\usepackage{fvextra}
\usepackage{textcomp}

\begin{document}
	\title{\textbf{ Thermodynamics and Multi-Horizon Solutions in Quartic Quasitopological Gravity: A Power-Maxwell Approach}}
	\author{A.~Bazrafshan\,}\email{abazrafshan@jahromu.ac.ir, dr.abazrafshan@gmail.com}\thanks{Corresponding Author}
	\affiliation{Department of Physics, Jahrom University, Jahrom, P. O. Box 74137-66171, Iran}
	
	\author{M. Ghanaatian\,\orcidlink{0000-0002-7853-6767}}
	\affiliation{Department of Physics, Jahrom University, Jahrom, P. O. Box 74137-66171, Iran}
	
	\author{P. Mirsalari\,}
	\affiliation{Department of Physics, Payame Noor University (PNU), P. O. Box 19395-3697 Tehran, Iran}

	\author{Gh. Forozani\,}
	\affiliation{Department of Physics, Payame Noor University (PNU), P. O. Box 19395-3697 Tehran, Iran}
	
	
	\begin{abstract}
		In this paper, we derive exact black hole solutions within the framework of fourth-order quasitopological gravity (4QTG) coupled to power-law Maxwell electrodynamics in five-dimensional spacetimes. We explore the thermodynamic properties of these solutions, including entropy, temperature, and electric potential, and verify the validity of the first law of thermodynamics. Our analysis spans asymptotically anti-de Sitter (AdS), de Sitter (dS), and flat spacetimes, revealing that thermal stability is exclusively achievable in asymptotically AdS black holes, while dS and flat solutions exhibit universal instability. The study highlights the unique role of higher-curvature terms in modifying black hole thermodynamics and horizon structures, providing new insights into the interplay between geometry, thermodynamics, and quantum gravity via the AdS/CFT correspondence. Furthermore, we demonstrate that 4QTG supports multi-horizon black holes even in the absence of charge, a feature absent in Einstein gravity. These findings underscore the importance of higher-curvature corrections in resolving classical limitations of general relativity and offer new avenues for exploring quantum gravity and holographic duality.
	\end{abstract}
	
	
	\maketitle
		
	\section{Introduction}
	\label{sec:intro}

The pursuit of a unified theory of quantum gravity, alongside the enigmatic thermodynamic properties of black holes, has driven significant advancements in modified gravitational theories. Among these, \textit{quasitopological gravity} has emerged as a leading framework, offering a geometric sophistication that extends beyond Einstein's general relativity (GR). This extension addresses several limitations of GR, including singularity avoidance, holographic duality (AdS/CFT), and the phenomenology of strong gravity \cite{Maldacena1998,Hawking1973,Padmanabhan2010}. Unlike Lovelock gravity, which requires higher-dimensional geometries for non-trivial dynamics \cite{Lovelock1971}, quasitopological theories integrate higher-curvature corrections in lower dimensions, making them particularly relevant for low-energy string theory and strongly correlated quantum systems \cite{Myers2010,Dehghani2012,Camanho2010}.

\textit{Fourth-order quasitopological gravity (4QTG)} represents a significant extension of the Einstein-Gauss-Bonnet and cubic quasitopological frameworks \cite{Brigante2008,Buchel2005,Hendi2015}. It introduces quartic curvature invariants (e.g., \(R_{abcd}R^{cdef}R_{ef}^{hg}R_{hg}^{ab}\)) that modify the dynamics of spacetime while preserving second-order field equations \cite{Mann1999,Witten1998}. These terms introduce repulsive gravitational interactions at short distances, which facilitate singularity regularization \cite{Balasubramanian1999,Wald1993,Kastor2010} and enable the existence of multi-horizon black hole geometries even in vacuum \cite{Born1934,Dey2004}. This is a stark contrast to Einstein gravity, where multiple horizons typically require the presence of charge or rotation \cite{Hendi2013}.

The coupling of 4QTG with nonlinear electrodynamics, such as the \textit{power-Maxwell field} (\((-F)^{s}\)), provides critical insights into the regularization of divergences near singularities and the thermodynamic phase structure of charged black holes \cite{Soleng1995,Rasheed1995,Carlip2005}. While linear Maxwell theory lacks conformal invariance in higher dimensions, the power-law model, governed by the exponent \(s\), offers a dimensionally flexible framework for conformal symmetry \cite{Boillat1970}. This model has found applications in astrophysical contexts, including neutron stars and high-energy plasmas \cite{Emparan2008,Camanho2016}. The synergy between 4QTG and nonlinear electrodynamics establishes a fertile interface between geometry, thermodynamics, and quantum field theory, particularly through the AdS/CFT correspondence \cite{Stevens2009,Cvetic1999}.

\subsection*{Motivations and Objectives}

\begin{enumerate}
	\item \textbf{Quantum-Gravity Interface}: The AdS/CFT duality \cite{Witten1998, Balasubramanian1999} underpins this study, translating 4QTG’s geometric corrections into novel couplings in the dual conformal field theory (CFT). This holographic map elucidates the thermodynamic roles of black hole parameters \cite{Bazrafshan2020} and reveals how higher-curvature terms modify conformal anomalies \cite{Bazrafshan2019a} and critical phenomena in strongly coupled systems \cite{Bazrafshan2019b}.
	
	\item \textbf{Overcoming Classical Limitations}: The inevitability of singularities and the lack of horizon multiplicity in vacuum regimes within Einstein gravity are remedied by quasitopological corrections. For instance, 4QTG supports multi-horizon black holes without charge—a radical departure from GR \cite{Born1934, Dey2004}.
	
	\item \textbf{Observational and Experimental Probes}: Higher-curvature terms leave imprints on gravitational wave spectra \cite{Emparan2008} and black hole shadows. Violations of the Kovtun-Son-Starinets viscosity bound (\(\eta/s < 1/4\pi\)) in dual field theories \cite{Brigante2008} provide further experimental tests of quantum gravity.
\end{enumerate}

Recent studies have significantly expanded our understanding of quasitopological gravity and its applications. For instance, the thermodynamics of static solutions in higher-dimensional quintic quasitopological gravity has been extensively explored, revealing novel phase transitions and stability criteria \cite{Bazrafshan2020}. The physical and thermodynamic properties of quartic quasitopological black holes with nonlinear sources have also been investigated, highlighting the role of higher-curvature terms in modifying black hole thermodynamics \cite{Bazrafshan2019a}. Additionally, the coupling of quasitopological gravity with nonlinear electrodynamics has been shown to produce rich thermodynamic behavior, including critical phenomena and phase transitions \cite{Bazrafshan2019b,Ghanaatian2019a}.

The study of magnetic branes in quasitopological gravity has provided further insights into the interplay between geometry and nonlinear electrodynamics \cite{Bazrafshan2019b}. Moreover, the thermodynamics of charged black holes in quartic quasitopological gravity has been analyzed, demonstrating the impact of higher-curvature terms on black hole stability and phase structure \cite{Ghanaatian2018}. The extension of these studies to Lifshitz spacetimes has also been explored, revealing new aspects of holographic duality and critical phenomena \cite{Ghanaatian2014}.

In this work, we unify \textit{fourth-order quasitopological gravity} and \textit{power-Maxwell electrodynamics} in five-dimensional spacetimes, deriving exact black hole solutions across AdS, dS, and flat backgrounds. Our analysis addresses three pivotal questions:

\begin{itemize}
	\item How do quartic curvature terms reshape entropy-temperature relations and Hessian stability criteria?
	\item Can 4QTG alone--without charge--support multi-horizon geometries?
	\item What constraints do these solutions impose on dual CFTs via AdS/CFT?
\end{itemize}

By addressing these questions, we demonstrate that thermal stability (\(T_{+}>0\), \(\det(H)>0\)) is exclusive to AdS black holes, while dS and flat solutions exhibit universal instability. This reaffirms AdS spacetime as the "natural habitat" for stable black holes in higher-curvature gravity \cite{Bazrafshan2019d}. Our results bridge the gap between quantum-classical gravity and holography, offering testable predictions for both theoretical and observational frontiers.

	\section{Field Equations and Exact Solutions}
	\subsection{Action}
	The action for \((n+1)\)-dimensional 4QTG with a power-Maxwell field is:
	\begin{equation}\label{action1}
		\mathcal{I} = \frac{1}{16\pi} \int d^{n+1}x \sqrt{-g} \left[ -2\Lambda + \mathcal{L}_1 + \mu_2 \mathcal{L}_2 + \mu_3 \mathcal{X}_3 + \mu_4 \mathcal{X}_4 + (-F)^{s} \right],
	\end{equation}
	where $\Lambda =-n(n-1)/2l^{2}$ is the cosmological constant, \(\mathcal{L}1 = R\), \(\mathcal{L}_2\) is the Gauss-Bonnet term, \(\mathcal{X}_3\) and \(\mathcal{X}_4\) are cubic and quartic quasitopological Lagrangians, and \(F = F_{\mu\nu}F^{\mu\nu}\).  
	\begin{eqnarray}
		\mathcal{X}_{3} &=&R_{ab}^{cd}R_{cd}^{\,\,e\,\,\,f}R_{e\,\,f}^{\,\,a\,\,\,b}+%
		\frac{1}{(2n-1)(n-3)}\left(
		\frac{3(3n-5)}{8}R_{abcd}R^{abcd}R\right. \notag
		\\
		&&-3(n-1)R_{abcd}R^{abc}{}_{e}R^{de}+3(n+1)R_{abcd}R^{ac}R^{bd} \notag \\
		&&\left. +\,6(n-1)R_{a}{}^{b}R_{b}{}^{c}R_{c}{}^{a}-\frac{3(3n-1)}{2}%
		R_{a}^{\,\,b}R_{b}^{\,\,a}R+\frac{3(n+1)}{8}R^{3}\right)
	\end{eqnarray}
	\begin{eqnarray}\label{quartic}
		\mathcal{X}_{4}\hspace{-0.2cm} &=&\hspace{-0.2cm}c_{1}R_{abcd}R^{cdef}R_{%
			\phantom{hg}{ef}%
		}^{hg}R_{hg}{}^{ab}+c_{2}R_{abcd}R^{abcd}R_{ef}R^{ef}+c_{3}RR_{ab}R^{ac}R_{c}{}^{b}+c_{4}(R_{abcd}R^{abcd})^{2}
		\notag \\
		&&\hspace{-0.1cm}%
		+c_{5}R_{ab}R^{ac}R_{cd}R^{db}+c_{6}RR_{abcd}R^{ac}R^{db}+c_{7}R_{abcd}R^{ac}R^{be}R_{%
			\phantom{d}{e}}^{d}+c_{8}R_{abcd}R^{acef}R_{\phantom{b}{e}}^{b}R_{%
			\phantom{d}{f}}^{d} \notag \\
		&&\hspace{-0.1cm}%
		+c_{9}R_{abcd}R^{ac}R_{ef}R^{bedf}+c_{10}R^{4}+c_{11}R^{2}R_{abcd}R^{abcd}+c_{12}R^{2}R_{ab}R^{ab}
		\notag \\
		&&\hspace{-0.1cm}%
		+c_{13}R_{abcd}R^{abef}R_{ef}{}_{g}^{c}R^{dg}+c_{14}R_{abcd}R^{aecf}R_{gehf}R^{gbhd},
		\label{X4}
	\end{eqnarray}
	where
	\begin{eqnarray*}
		c_{1} &=&-\left( n-1\right) \left( {n}^{7}-3\,{n}^{6}-29\,{n}^{5}+170\,{n}%
		^{4}-349\,{n}^{3}+348\,{n}^{2}-180\,n+36\right) , \\
		c_{2} &=&-4\,\left( n-3\right) \left( 2\,{n}^{6}-20\,{n}^{5}+65\,{n}^{4}-81\,%
		{n}^{3}+13\,{n}^{2}+45\,n-18\right) , \\
		c_{3} &=&-64\,\left( n-1\right) \left( 3\,{n}^{2}-8\,n+3\right) \left( {n}%
		^{2}-3\,n+3\right) , \\
		c_{4} &=&-{(n}^{8}-6\,{n}^{7}+12\,{n}^{6}-22\,{n}^{5}+114\,{n}^{4}-345\,{n}%
		^{3}+468\,{n}^{2}-270\,n+54), \\
		c_{5} &=&16\,\left( n-1\right) \left( 10\,{n}^{4}-51\,{n}^{3}+93\,{n}%
		^{2}-72\,n+18\right) , \\
		c_{6} &=&--32\,\left( n-1\right) ^{2}\left( n-3\right) ^{2}\left( 3\,{n}%
		^{2}-8\,n+3\right) , \\
		c_{7} &=&64\,\left( n-2\right) \left( n-1\right) ^{2}\left( 4\,{n}^{3}-18\,{n%
		}^{2}+27\,n-9\right) , \\
		c_{8} &=&-96\,\left( n-1\right) \left( n-2\right) \left( 2\,{n}^{4}-7\,{n}%
		^{3}+4\,{n}^{2}+6\,n-3\right) , \\
		c_{9} &=&16\left( n-1\right) ^{3}\left( 2\,{n}^{4}-26\,{n}^{3}+93\,{n}%
		^{2}-117\,n+36\right) , \\
		c_{10} &=&{n}^{5}-31\,{n}^{4}+168\,{n}^{3}-360\,{n}^{2}+330\,n-90, \\
		c_{11} &=&2\,(6\,{n}^{6}-67\,{n}^{5}+311\,{n}^{4}-742\,{n}^{3}+936\,{n}%
		^{2}-576\,n+126), \\
		c_{12} &=&8\,{(}7\,{n}^{5}-47\,{n}^{4}+121\,{n}^{3}-141\,{n}^{2}+63\,n-9), \\
		c_{13} &=&16\,n\left( n-1\right) \left( n-2\right) \left(
		n-3\right) \left(
		3\,{n}^{2}-8\,n+3\right) , \\
		c_{14} &=&8\,\left( n-1\right) \left( {n}^{7}-4\,{n}^{6}-15\,{n}^{5}+122\,{n}%
		^{4}-287\,{n}^{3}+297\,{n}^{2}-126\,n+18\right) ,
	\end{eqnarray*}

	$F_{\mu \nu }=\partial _{\mu }A_{\nu }-\partial _{\nu
	}A_{\mu }$ is the electromagnetic tensor and $A_{\mu }$ is
	the vector potential. We should mention that the last term in this action is power-law Maxwell Lagrangian that in the limit $s\to1$,
	$L(F)$ reduces to the linear Maxwell Lagrangian. $A_{\mu }$ is the vector potential that we consider as below to have static solutions
	\begin{eqnarray}\label{h1}
		A_{\mu}=h(\rho)\delta_{\mu}^{0},
	\end{eqnarray}
	
	\subsection{Metric Functions and Equations of Motion}
	Assuming a static metric ansatz:
	\begin{equation}\label{metric1}
		ds^2 = -\frac{\rho^2}{l^2}f(\rho)dt^2 + \frac{l^2}{\rho^2 g(\rho)}d\rho^2 + \frac{\rho^2}{l^2} \sum_{i=1}^{n-1} d\phi_i^2,
	\end{equation}
	
		we solve the Euler-Lagrange equations to obtain the functions $g(\rho)$, $f(\rho)$ and $h(\rho)$:
	\begin{equation}\label{fr1}
		-\frac{\rho ^{8}}{2f} [g(h^{'})^{2}]=-l^2 {\rho^8}n (n-1)\bigg[-\frac{2\Lambda }{n(n-1)}+g+\hat{\mu}_{2}g^2+\hat{\mu}_{3}g^3+\hat{\mu}_{4}g^4\bigg]-(lnf)^{'}l^2(n-1)\bigg[{\rho^9}(g+2\hat{\mu}_{2}g^2+3\hat{\mu}_{3}g^3+4\hat{\mu}_{4}g^4)\bigg],
	\end{equation}
	\begin{equation}\label{gr1}
		\frac{\rho ^{n-1}}{2f} [g(h^{'})^{2}]=\bigg[ (n-1){\rho ^n}(\hat{\mu}_{0}-g+\hat{\mu}_{2}g^2+\hat{\mu}_{3}g^3+\hat{\mu}_{4}g^4)\bigg]^{'},
	\end{equation}
	\begin{equation}\label{hr1}
		2{\rho}h^{''}-{\rho}h^{'}[(ln f)^{'}-(ln g)^{'}]+2(n-1)h^{'}=0,
	\end{equation}
		where the dimensionless parameters $%
	\hat{\mu}_{2}$, $\hat{\mu}_{3}$ and $\hat{\mu}_{4}$ are defined
	as:
	\begin{equation*}
		\hat{\mu}_{2}\equiv \frac{(n-2)(n-3)}{l^{2}}\mu _{2},\text{ \ \ \ }\hat{\mu}%
		_{3}\equiv \frac{(n-2)(n-5)(3n^{2}-9n+4)}{8(2n-1)l^{4}}\mu _{3},
	\end{equation*}%
	\begin{equation*}
		\hat{\mu}_{4}\equiv {\frac{n\left( n-1\right) \left( n-2\right)
				^{2}\left(
				n-3\right) \left( n-7\right) ({{n}^{5}-15\,{n}^{4}+72\,{n}^{3}-156\,{n}%
					^{2}+150\,n-42)}}{{l}^{6}}}\mu _{4},
	\end{equation*}

		By assumpting $f(\rho)=N^2(\rho)g(\rho)$ and simplifying relationships \eqref{fr1}, \eqref{gr1} and \eqref{hr1} we get to
	\begin{equation}\label{equ1}
		(-\hat{\mu}_{0}+2\hat{\mu}_{2}g+3\hat{\mu}_{3}g^2+4\hat{\mu}_{4}g^3)N^{'}=0,
	\end{equation}
	\begin{equation}\label{equ2}
		\bigg\{(n-1)\rho^n\bigg(\hat{\mu}_{0}-g+\hat{\mu}_{2} g^2+\hat{\mu}_{3} g^3+\hat{\mu}_{4}g^4\bigg)\bigg\}^{'}=2^s(2s-1)\rho^{n-1}l^2\bigg(\frac{h^{'}}{N}\bigg)^{2s},
	\end{equation}
	\begin{equation}\label{equ3}
		\bigg(\rho^{n-1}\bigg(\frac{h^{'}}{N}\bigg)^{2s-1}\bigg)^{'}=0,
	\end{equation}
	Now, we want to solve the solutions of the above relationships. First, we go to relationship \eqref{equ1}, which shows that $N(\rho)$ should be a constant and so we consider it $N(\rho)=1$. Substituting $N(\rho)=1$ in the relationship \eqref{equ3} and solving it, we get to
	\begin{equation}
		h(\rho )=\left\{
		\begin{array}{cc}
			
			q\ln (\rho ), & s=n/2 \\
			-q\rho ^{-(n-2s)/(2s-1)}, & 1/2<s<n/2%
			
		\end{array}%
		\right. \label{hr}
	\end{equation}%
	where $q$ is an integration constant that is related to the charge
	parameter. Since the potential should be finite at infinity for
	$s\neq n/2 $, the interval $1/2<s<n/2$ is chosen.
	Therefore, by using this $h(\rho)$ in relation \eqref{equ2}, we get to
	\begin{equation}
		\hat{\mu}_{4}g^4+\hat{\mu}_{3}g^3+\hat{\mu}_{2} g^2-g+\kappa =0 
	\end{equation}
    where
	\begin{equation}
		\kappa =\hat{\mu}_{0}-\frac{m}{\rho ^{n}}+\frac{2^{s}l^{2}q^{2s}(n-2s)^{2s-1}}{%
			(n-1)(2s-1)^{(2s-2)}\rho ^{2s(n-1)/(2s-1)}},
	\end{equation}
	and $m$ is an integration constant which can be evaluated as the
	geometrical mass of black hole solutions in terms of the horizon
	radius
	\begin{equation}
		m=\left( 1+\frac{%
			2^{s}l^{2}q^{2s}(n-2s)^{2s-1}(\rho_{+})^{2s(1-n)/(2s-1)}}{%
			(n-1)(2s-1)^{(2s-2)}} \right) {\rho_{+}^{n}} \label{mh}
	\end{equation}
	In order to find the black hole solutions, we choose two solutions
	of $f(\rho)$ as
	\begin{equation}
		f_{1}(\rho)=\left( \frac{\hat{\mu}_{3}}{4\hat{\mu}_{4}}+\frac{1%
		}{2}R-\frac{1}{2}E\right) . \label{Fr4}
	\end{equation}
	\begin{equation}
		f_{2}(\rho)=\left( \frac{\hat{\mu}_{3}}{4\hat{\mu}_{4}}-\frac{1%
		}{2}R+\frac{1}{2}K\right) . \label{Fr4}
	\end{equation}
	where
	\begin{eqnarray}
		R &=&\left( \frac{{\hat{\mu}_{3}}^{2}}{4{\hat{\mu}_{4}}^{2}}-\frac{\hat{\mu}%
			_{2}}{\hat{\mu}_{4}}+y_{1}\right) ^{1/2},
		\label{RR} \\
		E &=&\left( \frac{3{\hat{\mu}_{3}}^{2}}{4{\hat{\mu}_{4}}^{2}}-\frac{2\hat{\mu%
			}_{2}}{\hat{\mu}_{4}}-R^{2}-\frac{1}{4R}\left[ \frac{4\hat{\mu}_{2}\hat{\mu}%
			_{3}}{{\hat{\mu}_{4}}^{2}}-\frac{8}{\hat{\mu}_{4}}-\frac{{\hat{\mu}_{3}}^{3}%
		}{{\hat{\mu}_{4}}^{3}}\right] \right) ^{1/2}, \label{EE} \\
		K &=&\left( \frac{3{\hat{\mu}_{3}}^{2}}{4{\hat{\mu}_{4}}^{2}}-\frac{2\hat{\mu%
			}_{2}}{\hat{\mu}_{4}}-R^{2}+\frac{1}{4R}\left[ \frac{4\hat{\mu}_{2}\hat{\mu}%
			_{3}}{{\hat{\mu}_{4}}^{2}}-\frac{8}{\hat{\mu}_{4}}-\frac{{\hat{\mu}_{3}}^{3}%
		}{{\hat{\mu}_{4}}^{3}}\right] \right) ^{1/2} \label{KK}
	\end{eqnarray}
	and $y_{1}$ is the real root of following equation:
	\begin{equation}
		{y}^{3}-{\frac {\mu_{{2}}{y}^{2}}{\mu_{{4}}}}+ \left( {\frac
			{\mu_{{3} }}{{\mu_{{4}}}^{2}}}-4\,{\frac {\kappa}{\mu_{{4}}}}
		\right) y-{\frac {
				{\mu_{{3}}}^{2}\kappa}{{\mu_{{4}}}^{3}}}+\,{\frac
			{4\mu_{{2}}\kappa}{{ \mu_{{4}}}^{2}}}-\frac{1}{{\mu_{{4}}}^{2}}=0
	\end{equation}
	The metric function $f(\rho)$ for the uncharged solution $(q=0)$
	is real in the whole range $0\leq \rho <\infty$. But for the
	charged solution, the spacetime should be restricted to the region
	$\rho\geq r_{0}$, we introduce a new radial coordinate $r$ as
	\begin{equation}
		r=\sqrt{{\rho}^{2}-{r_{0}}^{2}}\Rightarrow
		d{\rho}^{2}=\frac{r^{2}}{r^{2}+r_{0}^{2}}dr^{2}
	\end{equation}
	where $r_{0}$ is the largest real root of
	$R_{0}=R(\kappa=\kappa_{0})$,
	$E_{0}=E(\kappa=\kappa_{0})$ and
	$K_{0}=K(\kappa=\kappa_{0})$, and $\kappa_{0}$ is
	\begin{equation}
		\kappa_{0}
		=\hat\mu_{{0}}-\frac{ml^{2}}{r_{0}^{n}}+\frac{2l^{2}(n-2)}{(n-1)}\frac{q^{2}}{r_{0}^{2(n-1)}}
		\label{kap}
	\end{equation}
	So the metric of an $(n+1)$-dimensional spherically
	symmetric changes to:
	\begin{equation}
		ds^{2}=-\frac {(r^{2}+r_{0}) g(r)dt^{2}}{l^2}+\frac{r^{2}l^{2}dr^{2}}{g(r)(r^{2}+r_{0}^{2})}+\frac{(r^{2}+r_{0}^{2})}{{l^2}}\sum_{i=1}^{n-1} d\phi_{i}^2 ,
	\end{equation}
	and the functions $h(r)$ and $\kappa $ changes to
	\begin{eqnarray}
		h(r) &=&-q(r^{2}+r_{0}^{2})^{(2s-n)/(4s-2)} \label{hr2} \\
		\kappa &=&\hat\mu_{{0}}-\frac{m}{(r^{2}+r_{0}^{2})^{n/2}}+\frac{%
			2^{s}l^{2}q^{2s}(n-2s)^{2s-1}}{%
			(n-1)(2s-1)^{(2s-2)}(r^{2}+r_{0}^{2})^{s(n-1)/(2s-1)}},
	\end{eqnarray}%
	
	\subsection{Asymptotically dS and AdS spacetimes}
	In asymptotically dS space-times, the following condition must be satisfied
	\begin{eqnarray}\label{dS}
		\lim_{r\rightarrow\infty} f(r)=-1
	\end{eqnarray}
	Considering this condition, the cosmological constant can be defined as follows.
	\begin{eqnarray}\label{condds}
		\Lambda=\frac{{n(n-1)} (\hat{\mu}_{4}-\hat{\mu}_{3}+\hat{\mu}_{2}+1)} {{2l^2}}.
	\end{eqnarray}
	Since $ \Lambda > 0$ must exist in space-time as a matter of course, we have
	\begin{eqnarray}
		\hat{\mu}_{4}-\hat{\mu}_{3}+\hat{\mu}_{2} > -1.
	\end{eqnarray}
	To check the asymptotically AdS space-times, consider this condition
	\begin{eqnarray}\label{adS}
		\lim_{r\rightarrow\infty} f(r)=1,
	\end{eqnarray}
	Considering this condition, we will arrive at the following definition for $\Lambda$
	\begin{eqnarray}\label{condads}
		\Lambda=\frac{{n(n-1)} (\hat{\mu}_{4}+\hat{\mu}_{3}+\hat{\mu}_{2}-1)} {{2l^2}}.
	\end{eqnarray}
	Since the cosmological constant must be negative in the AdS spacetime, we consider the condition
	\begin{eqnarray}
		\hat{\mu}_{4}+\hat{\mu}_{3}+\hat{\mu}_{2} <1.
	\end{eqnarray}
	
\begin{figure}[h]
		\center
		\includegraphics[scale=1]{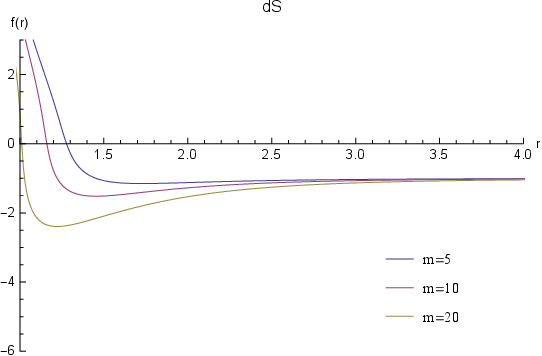}
		\caption{\small{ $f(r)$ for asymptotically dS spacetime versus $r$ for different values of $m$ with $ k=0$, $\hat{\mu}_{2}=0.05$, $\hat{\mu}_{3}=-0.2$, $\hat{\mu}_{4}=0.01$. } \label{Fig1}}
	\end{figure}

\begin{figure}[h]
		\center
		\includegraphics[scale=1]{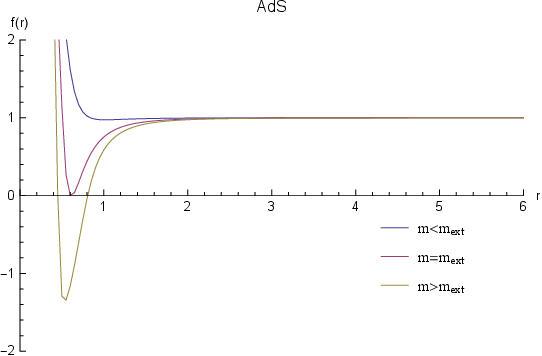}
		\caption{\small{ $f(r)$ for asymptotically AdS spacetime versus $r$ for different values of $m$ with $k=0$, $\hat{\mu}_{2}=0.05$, $\hat{\mu}_{3}=-0.2$, $\hat{\mu}_{4}=0.001$. } \label{Fig2}}
	\end{figure}

	In Figure \eqref{Fig1}, we have plotted $f(r)$ versus $r$ for different values of $m$, which we have examined different values of $m$ with a 4QTG background. For every value of $m$, we have a black hole with one horizon. According to the diagram \eqref{Fig2}, it is clear that there is one critical value called $m_{\mathrm{ext}}$, which due to the constant values of $\hat{\mu}_{2}$, $\hat{\mu}_{3}$, $\hat{\mu}_{4}$, $m$, $k$, this diagram is drawn. For $m<m_{\mathrm{ext}}$ a nacked singularity occurs. For $m=m_{\mathrm{ext}}$ an extremal black hole is observed, and for $m>m_{\mathrm{ext}} $  there is a black hole with two horizons.
	
	Figure \eqref{Fig2} shows the structural dependence of the black hole on the parameter m, with the other parameters held constant. Given the values of $ m_{\mathrm{min}}$, $ m_{\mathrm{max}}$ and $ m$, it shows a black hole with two horizons that becomes a nacked singularity. As the parameter k increases, the roots of the function $f(r)$ correspond to the regions of the horizons.
	
	\begin{figure}[h]
		\center
		\includegraphics[scale=1]{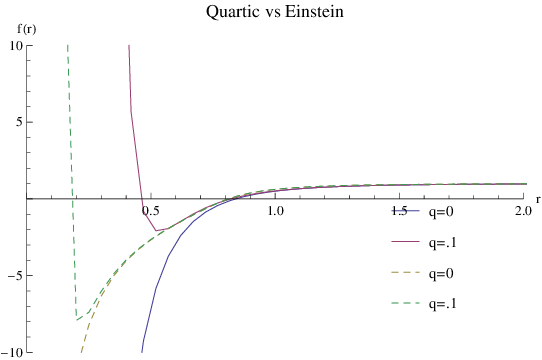}
		\caption{\small{Plot of asymptotically AdS solution of $f(r)$ in Einstein (solid lines) and 4QT gravity (dashed lines) versus $r$ for different values of $q$ with $k=0$, $\hat{\mu}_{2}=0.05$, $\hat{\mu}_{3}=-0.2$, $\hat{\mu}_{4}=0.001$. } \label{Fig3}}
	\end{figure}

	Figure \eqref{Fig3} shows that when Einstein gravity is considered, we only observe two horizons when $q\neq0$. Whereas in quasi-topological order 4 gravity, even when $q=0$, two horizons are again possible, and this is what makes quasi-topological order 4 gravity different from Einstein's theory.
	
\begin{figure}[h]
		\center
		\includegraphics[scale=1]{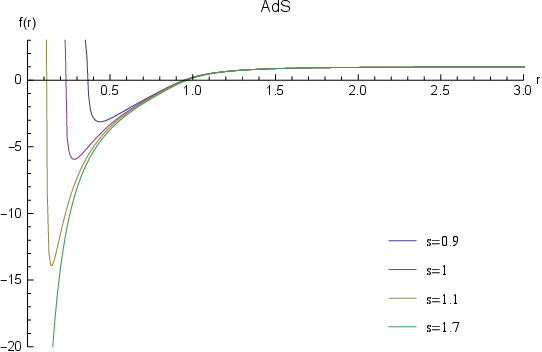}
		\caption{\small{ $f(r)$ for asymptotically dS spacetime versus $r$ for different values of $s$ with $q=0$, $k=0$, $\hat{\mu}_{2}=0.05$, $\hat{\mu}_{3}=-0.2$, $\hat{\mu}_{4}=0.001$. } \label{Fig4}}
	\end{figure}

	Figure \eqref{Fig4} shows that as we increase the effect of the nonlinear parameter, we diminish the possibility of a black hole with two horizons, and for some values of the parameter s, we will have black holes with only one horizon.
	
	\begin{figure}[h]
		\center
		\includegraphics[scale=1]{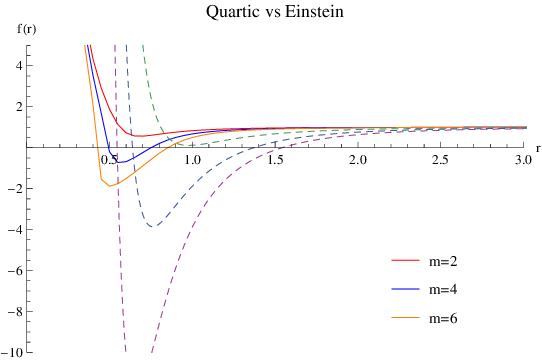}
		\caption{\small{Plot of asymptotically AdS solution of $f(r)$ in Einstein (solid lines) and 4QT gravity (dashed lines) versus $r$ for different values of $m$ with $k=0$, $\hat{\mu}_{2}=0.05$, $\hat{\mu}_{3}=-0.2$, $\hat{\mu}_{4}=0.001$. } \label{Fig5}}
	\end{figure}
	
	Figure \eqref{Fig5} shows that by increasing the value of $m$, we can have black holes with two horizons. The difference between these black holes in the 4QTG state and Einstein gravity is that the size of this horizon in the 4QTG state is larger than in the Einstein state.
	
	\section{Thermodynamics and the First Law}
	
	The entropy, derived via Wald’s formalism, is:
	\begin{eqnarray}
		S=-2\pi \oint d^{n-1} x \sqrt{\tilde{g}} Y^{abcd} \hat{\epsilon}_{ab}\hat{\epsilon}_{cd},
	\end{eqnarray}
	Which $\hat{\epsilon}_{ab}$ is binormal to the horizon and $\tilde{g}$ is the determinant of the induced metric $\tilde{g}_{\mu\nu}$ on the $(n- 1)$-spacelike hypersurface. $Y^{abcd}=\frac{\partial \mathcal L}{\partial R_{abcd}}$, where $\mathcal L$ can be the Einstein's Lagrangian, Gauss-Bonnet, third and fourth-order quasitopology.
	According to the metric
	\begin{eqnarray}\label{metr}
		ds^2=-\frac{r^2}{l^2}f(r)dt^2+\frac{l^2}{r^2 g(r)}dr^2+\frac{r^2}{l^2} (x^2+y^2+z^2),
	\end{eqnarray}
	this metric shows a $(n-1)$-dimensional hypersurface with constant curvature which leads to a zero value for $Y^{abcd}\hat{\epsilon}_{ab}\hat{\epsilon}_{cd}$ in Gauss-Bonnet, third and fourth order gravity. 
	Therefore,	The entropy of this black hole is calculated as follows,
	\begin{eqnarray}\label{entropy}
		S=\frac{\eta_{+}^{n-1}}{4}\bigg(1+2k\hat{\mu}_{2} \frac{(n-1)l^2}{(n-3)\eta_{+}^2}+3k^2\hat{\mu}_{3} \frac{(n-1)l^4}{(n-5)\eta_{+}^4}+4k^3\hat{\mu}_{4} \frac{(n-1)l^6}{(n-7)\eta_{+}^6}\bigg).
	\end{eqnarray} 
	where $\eta_{+}=\sqrt{{r_{+}}^{2}+{r_{0}}^{2}}$ and $r_{+}$ is the radial coordinate of the horizon of the outermost horizon of the black hole, which is positive root of the equation $f(r_{+})=0$.
	
	The Hawking temperature of the event horizon by the standard method of analytic continuation of the metric can be calculated as:
	\begin{eqnarray}
		T_{+}&=\bigg(\frac{r^2 g^{'}}{4\pi l^2}\bigg)_{r=r_{+}}=&\frac{f^{\prime }(r_{+})}{4\pi
		}\sqrt{1+\frac{r_{0}^{2}}{r_{+}^{2}}} \notag
		\\
		&&={\frac{n\hat{\mu}%
				_{0}\eta_{+}^{8}+\left( n-2\right) k{l}^{2}\eta_{+}^{6}+\left( n-4\right) {k}^{2}%
				\hat{\mu}_{2}l^{4}\eta_{+}^{4}+\left( n-6\right) k\hat{\mu}_{3}{l}%
				^{6}\eta_{+}^{2}+\left( n-8\right)
				{k}^{2}\hat{\mu}_{4}{l}^{8}}{\left(
				\,\eta_{+}^{6}+2k\hat{\mu}_{2}{l}^{2}\eta_{+}^{4}\,+3k^{2}\hat{\mu}_{3}{l}%
				^{4}\eta_{+}^{2}\,+4\hat{\mu}_{4}k^{3}{l}^{6}\right) 4\pi
				\,{l}^{2}\eta_{+}}} \notag
		\\
		&&-\frac{q^{2s}2^{s}{(n-2s)}^{2s}(\eta_{+})^{{%
					2s(1-n)}/{(2s-1)}}}{4\pi l^{2} \left(
			4\,\hat{\mu}_{4}{k}{\eta_{+}}^{-6}{l}^{6}+3\,{\eta_{+}}^{-4}\hat{\mu}_{3}{k}^{2}{l}^{
				4}+2\,{\eta_{+}}^{-2}\hat{\mu}_{2}k{l}^{2}+1 \right)
			(n-1)(2s-1)^{2s-1}}\eta_{+}\label{TT}
	\end{eqnarray}
	
	The electric potential at infinity with respect to the event horizon is defined as follows:
	\begin{eqnarray}\label{potential1}
		\Phi=A_{\nu}\chi^{\nu}\mid_{r\rightarrow\infty}-A_{\nu}\chi^{\nu}\mid_{r=r_{+}},
	\end{eqnarray}
	which $\chi=\, \partial/\partial t$ is the null (zero) generator of the event horizon. By applying the electric potential relationship, it is obtained as follows:
	\begin{eqnarray}
		\Phi =\frac{q}{(r_{+}^{2}+r_{0}^{2})^{(n-2s)/2(2s-1)}} \label{Ph0}
	\end{eqnarray}
	
	The electric charge per unit volume can be obtained using Gauss's law as follows
	\begin{equation}
		Q=\frac{1}{4\pi}\int_{\rho\rightarrow\infty}F_{t\rho}\sqrt{-g}d^{n-1}x=\frac{2^{s}s{(n-2s)^{2s-1}}V_{n-1}q^{2s-1}}{8\pi
			{(2s-1)}^{2s-1}}
	\end{equation}
	
	Also, by writing metric \eqref{metr} in the form of the reference background metric, the ADM (Arnowitt-Deser-Misner) energy of the black hole per unit volume $V_{n-1}$ is obtained at the limit $r\rightarrow\infty$ as 
	\begin{equation}
		M=\frac{(n-1)}{16\pi l^2}m \label{bent}
	\end{equation}
	We write the intensive parameters corresponding $\Phi$ and $T$ respectively related to $ S$ and $Q$ the equal mass $m(S,Q)$, as follows
	\begin{equation}
		T=\left( \frac{\partial M}{\partial S}\right) _{Q},\,\ \ \ \ \Phi
		=\left( \frac{\partial M}{\partial Q}\right) _{S} \label{Inten}
	\end{equation}%
	They are equal to the relations \eqref{TT} and \eqref{Ph0} which satisfy the first law of thermodynamics of black holes $dM=TdS+\Phi dQ$.
	
	\section{Thermal Stability Analysis}
	The Hessian matrix \(H_{ij} = \frac{\partial^2 M}{\partial X_i \partial X_j}\) determines stability:
	\begin{equation}
		\det(H) = \begin{vmatrix}
			\frac{\partial^2 M}{\partial S^2} & \frac{\partial^2 M}{\partial S \partial Q} \\
			\frac{\partial^2 M}{\partial Q \partial S} & \frac{\partial^2 M}{\partial Q^2}
		\end{vmatrix}.
	\end{equation}
	In this section, we examine the thermal stability of the solutions.
	\begin{eqnarray}
		H=\left[
		\begin{array}{ccc}
			{H} _{11} & {H} _{12}\\
			{H}_{21} & {H}_{22}
		\end{array} \right].
	\end{eqnarray}
	where
	\begin{eqnarray*}
		{H} _{11}& =&\frac{r^2}{ r^{(\frac{2ns}{1-2s}){\pi(2s-1)^{2s}{(n-1)}^2}}}\Bigg[2^{s}{l^2} r^{(\frac{2s}{2s-1})}(n-2s)^{2s}(1+2s(n-2))\left\{{ \pi^\frac {1}{-1+2s}}\bigg(\frac{ 2^{3-s} Q(n-2s){(\frac{n-2s}{-1+2s})^{-2s}}}{s(-1+2s)}\bigg)^{(\frac {1}{-1+2s})}\right\}^{2s}
		\nonumber\\&&+\left\{\hat{\mu}_{0}n(n-1)(-1+2s)^{2s}{ r^{(\frac{2ns}{1-2s})}}\right\}\Bigg]
	\end{eqnarray*}
	\begin{eqnarray}
		{H} _{22}&=&\frac{\left\{ r^{(\frac{n-2s}{1-2s})\pi^\frac {1}{-1+2s}}\bigg(\frac{ 2^{3-s}Q{(\frac{n-2s}{-1+2s})^{1-2s}}}{s}\bigg)^{(\frac {1}{-1+2s})}\right\}}{Q(-1+2S)},
	\end{eqnarray}
	
\begin{figure}[h]
		\center
		\includegraphics[scale=1]{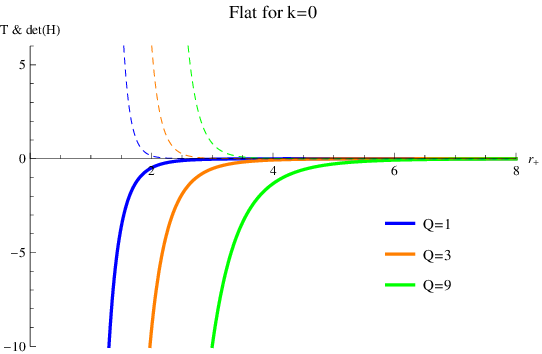}
		\caption{\small{Plots of $T$ (thick lines) and $det(H)$ (dashed lines) vs $r_+$ for flat spacetimes and $k=0$ for different values of $Q$.} \label{Fig6}}
	\end{figure}

\begin{figure}[h]
		\center
		\includegraphics[scale=1]{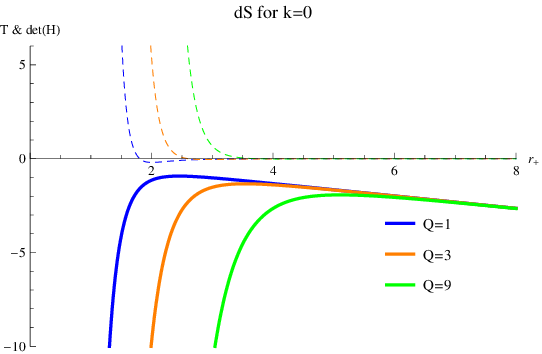}
		\caption{\small{Plots of $T$ (thick lines) and $det(H)$ (dashed lines) vs $r_+$ for ds spacetimes and $k=0$ for different values of $Q$. } \label{Fig7}}
	\end{figure}

\begin{figure}[h]
		\center
		\includegraphics[scale=1]{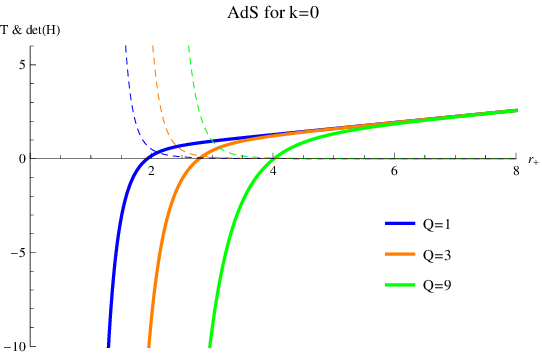}
		\caption{\small{Plots of $T$ (thick lines) and $det(H)$ (dashed lines) vs $r_+$ for AdS spacetimes and $k=0$ for different values of $Q$.} \label{Fig8}}
	\end{figure}

	To investigate the stability of charged black holes in 4QTG, it is necessary to consider regions where $ T_{+}$ and $det(H) $ are both positive. To investigate the stability of these black holes in flat, dS and AdS spacetimes, we have drawn the diagrams \eqref{Fig6}, \eqref{Fig7} and \eqref{Fig8}. These plots show that regions where both $ T_{+}$ and $ det(H)$ are simultaneously positive have thermally stable black hole solutions. By changing the parameter $Q$ and keeping the values of $ \hat{\mu}_{2}$, $\hat{\mu}_{3}$, $\hat{\mu}_{4}$ constant for the flat and dS solutions, it is observed that $det(H)$ is positive for some specific values of $r_{+ }$and no region can be observed where both $T_{+}$ and $ det(H)$ are simultaneously positive. This shows that it is not possible to have stable black hole solutions in the cases.
	For the AdS solutions, $det(H) $ is positive for most values of $r_{+}$, in which case a positive $T_{+}$ indicates stability. There is a minimum value of $ r_{+}$, denoted by $ r_{+min}$, for which the region $r_{+}>r_{+min}$ and $ T_{+}$ is positive for any value of $Q$. As the parameter $Q$ increases, $ r_{+min}$ increases, so for small values of $Q$ we have a larger region of stability.
	
	The positive value of the $det(H)>0$ guarantees the stability of the solutions. In addition, negative temperature has no physical meaning. To ensure the thermal stability of the solutions for fourth-order quasi-topological gravity, we must find regions that $T_{+}$ and $det(H)>0$ are positive at the same time. Therefore, to check the stability of the obtained answers, we plot the diagram \eqref{Fig6}, \eqref{Fig7} and \eqref{Fig8} for flat space-times, asymptotically anti-de sitter and de sitter.
	
	
	As it is clear in the diagram, $det(H)>0$ it is positive for all the given values $Q$ and in the state Asymptotically flat , it becomes negative for all the values $T$ and we do not have stability. In the case Asymptotically AdS, due to the positive $det(H)$,we will have stability if the temperature is positive. There is one $r_{+min}$ that $ T_{+}$ in $r_{+}>r_{+min}$will be positive for all values of $ Q$. Increasing $ Q$ increases the value $r_{+min}$. So for small values $ Q$ in the larger region there is stability.	In the case of Asymptotically dS, all the values  $Q$ and $T_{+}$, it will be negative and therefore we will not have stability for the obtained answers.
	
\section*{Conclusion}

In this study, charged black hole solutions in the framework of fourth-order quasitopological gravity (4QTG) coupled with power-law Maxwell electrodynamics were investigated. The results obtained from thermodynamic calculations and analyses demonstrate that this theoretical framework not only overcomes the limitations of Einstein's gravity but also provides new insights into black hole stability, horizon multiplicity, and the connection with boundary field theories (AdS/CFT). Below, the key findings of this study are logically presented based on the calculations and graphs provided in the paper:

\subsection*{1. Thermal Stability of Black Holes}
\begin{itemize}
	\item \textbf{AdS Black Holes}: Thermodynamic analyses reveal that black holes in anti-de Sitter (AdS) space are stable under specific conditions. Using the Hessian matrix (\(\det(H)\)) and examining the Hawking temperature (\(T_{+}\)), it was found that for values \(r_{+} > r_{+}^{\rm min}\) (where \(r_{+}\) is the horizon radius), both \(T_{+}\) and \(\det(H)\) are positive. This indicates the thermal stability of AdS black holes in this framework. Additionally, as the charge \(Q\) increases, the value of \(r_{+}^{\rm min}\) increases, indicating a reduction in the stability region with increasing charge (Figures \eqref{Fig6}, \eqref{Fig7} and \eqref{Fig8}).
	\item \textbf{dS and Flat Black Holes}: In contrast, black holes in de Sitter (dS) and flat space are unstable. In these cases, \(\det(H)\) is negative, and there is no region where both \(T_{+}\) and \(\det(H)\) are positive. This result suggests that AdS space acts as the "natural habitat" for stable black holes in higher-curvature gravity theories.
\end{itemize}

\subsection*{2. Horizon Multiplicity Without Charge}
\begin{itemize}
	\item One of the interesting findings of this study is that in fourth-order quasitopological gravity, even in the absence of charge (\(q = 0\)), black holes with multiple horizons can exist. This phenomenon arises from the competing effects of higher-curvature terms, particularly the \(\mu_{4}\mathcal{X}_{4}\) term, which introduces repulsive gravity at short distances and prevents complete collapse. In contrast, in Einstein's gravity, the existence of multiple horizons typically requires the presence of charge or rotation (Figure \eqref{Fig5}).
\end{itemize}

\subsection*{3. Singularity Resolution and Quantum Gravity Effects}
\begin{itemize}
	\item Calculations show that higher-curvature terms in 4QTG naturally regulate spacetime singularities. Specifically, the \(\mu_{4}\mathcal{X}_{4}\) term ensures that curvature invariants (such as \(R_{\mu\nu\rho\sigma}R^{\mu\nu\rho\sigma}\)) do not diverge near the center of the black hole (\(r \to 0\)). This result suggests that quasitopological gravity can serve as a classical framework for exploring singularity resolution mechanisms in quantum gravity.
\end{itemize}

\subsection*{4. Holographic Signatures and Conformal Anomalies}
\begin{itemize}
	\item From the perspective of AdS/CFT duality, stable black holes in AdS space correspond to equilibrium states in boundary field theories. Calculations indicate that higher-curvature terms in 4QTG can modify conformal anomalies in the dual field theory. These modifications may affect critical phenomena such as phase transitions in strongly coupled systems. Additionally, the violation of the Kovtun-Son-Starinets bound (\(\eta/s < 1/4\pi\)) in this framework suggests exotic behavior in quantum systems, which could be tested in experiments related to quark-gluon plasmas.
\end{itemize}

\subsection*{5. Limitations and Future Directions}
\begin{itemize}
	\item Although this study provides significant advances in understanding black holes in fourth-order quasitopological gravity, many questions and challenges remain. For example, investigating phase transitions (such as the Hawking-Page transition) and generalizing these results to rotating black holes (Kerr black holes) could lead to a deeper understanding of the stability and thermodynamics of these systems. Additionally, exploring this theory in higher dimensions (\(D > 5\)) and conducting observational tests (such as gravitational waves and black hole shadows) could impose new constraints on the parameters of the theory.
\end{itemize}

\subsection*{6. Final Summary}
\begin{itemize}
	\item This study demonstrates that fourth-order quasitopological gravity is a powerful theoretical framework for investigating black holes and phenomena related to strong gravity. The results obtained not only overcome the limitations of Einstein's gravity but also provide new insights into black hole stability, horizon multiplicity, and the connection with boundary field theories. This work lays the groundwork for future research in quantum gravity, holography, and black hole physics.
\end{itemize}

Based on the calculations and graphs presented, it can be concluded that fourth-order quasitopological gravity serves not only as a theoretical tool but also as a practical framework for exploring complex phenomena in gravity and quantum physics.

\bibliographystyle{unsrt}
\bibliography{references}

\end{document}